\documentclass[reprint,aps,ams,amsmath,prx,twocolumn,longbibliography]{revtex4-1}
\usepackage{graphics}
\usepackage{graphicx}
\usepackage{epsfig}
\usepackage{amsmath}
\usepackage{amssymb}
\usepackage{amsfonts}
\usepackage{bm}
\usepackage{color}
\usepackage{bbm}
\usepackage{hyperref}
\usepackage[normalem]{ulem}
\usepackage{comment}
\usepackage{soul}

\begin{document}

\title{Spin-caloric resistance of Dirac plasma in a graphene Corbino device}

\author{Alex Levchenko}
\affiliation{Department of Physics, University of Wisconsin-Madison, Madison, Wisconsin 53706, USA}

\date{December 9, 2024}

\begin{abstract}
The thermal resistance of a spin-polarized hydrodynamic Dirac plasma in graphene is considered. A mechanism for the coupling of heat and spin flows is discussed, demonstrating that spin diffusion and spin thermocurrent modify viscous dissipation, leading to a significant enhancement of thermal resistance. Practical calculations are then presented for graphene devices in the Corbino geometry.
\end{abstract}

\maketitle

\section{Introduction}

Thermoelectric (TE) and thermomagnetic (TM) transport effects in metals reflect particle-hole asymmetry and arise from the coupling of charge and heat currents \cite{Abrikosov}. 
A different class of thermomagnetoelectric (TME) effects emerges when one considers coupling of charge and heat flow with the spin current \cite{Aronov:1976,Silsbee:1987}.  
The distinction between TM and TME phenomena lies in the former being driven by the Lorentz force of the applied external magnetic field exerted on the charge flow, 
while the latter arises from the flow of spins driven by a magnetic field.

In metallic magnetic heterostructures, such as spin valves, spin-dependent thermoelectric effects can be understood through two main models. The first model involves two parallel spin-transport channels with distinct thermoelectric properties. Alternatively, these effects can be attributed to collective phenomena, influenced by spin waves, which are also present in insulating ferromagnets. For further references, refer to the collection of papers and reviews on spin caloritronics \cite{Bauer:2010,Bauer:2012}.
The two-channel model, incorporating spin-dependent conductivities, can be derived from the kinetic equation. This model assumes different drift velocities for two spin subsystems and is particularly suitable in the weak interaction limit. Essentially, it extends the single-particle Drude-Sommerfeld theory to account for spin-polarized transport. Consequently, it adheres to Mott's law for the Seebeck coefficient and the Wiedemann-Franz law for spin-dependent thermal conductivity. In contrast, in the collision-dominated regime, the drift velocity is the same for all the spin components. The reason is that frequent collisions between electrons with different spins
form a common drift of the electron system. This scenario can be realized in high-mobility and low-density semiconductor and semimetal devices, where strong correlations lead to the electron liquid reaching a hydrodynamic limit \cite{Gurzhi,AKS}. In this situation, the coupling of spin and heat currents exhibits a distinct character, leading to the specific mechanism of thermal resistance explored in this paper.

The hydrodynamic behavior of the electron liquid has been demonstrated in a number of recent experiments, see reviews \cite{Lucas:2018,AL:2020,Narozhny:2022,Fritz:2024} on this topic and references therein. Of particular relevance to this work are strong thermoelectric anomalies observed in graphene devices tuned to the proximity of charge neutrality \cite{Crossno:2016,Ghahari:2016}. Physically, these profound features, namely giant enhancement of the Lorenz ratio and violation of the Mott relation by inelastic scattering, stem from the decoupling of charge and heat flows. 

The hydrodynamic equations in the presence of finite spin-polarization of an electron liquid induced by an in-plane magnetic field should be supplemented by the conservation law for spin. This is particularly relevant to graphene multilayers, as the spin relaxation time is known to be long, even at room temperature, in the range of nanoseconds, with corresponding spin diffusion lengths reaching 10 $\mu$m  \cite{Drogeler:2014,Drogeler:2016}. In the linear response, the spin current consists of a spin diffusion current and a spin thermocurrent induced by temperature gradients. According to Onsager's reciprocity principle \cite{Onsager:1931}, the heat flux acquires an additional spin-dependent contribution related to the Peltier effect. These processes modify the viscous flow of the electron fluid, thus influencing dissipative friction and consequently strongly increasing thermal resistance.

\begin{figure}[t!]
\includegraphics[width=0.9\linewidth]{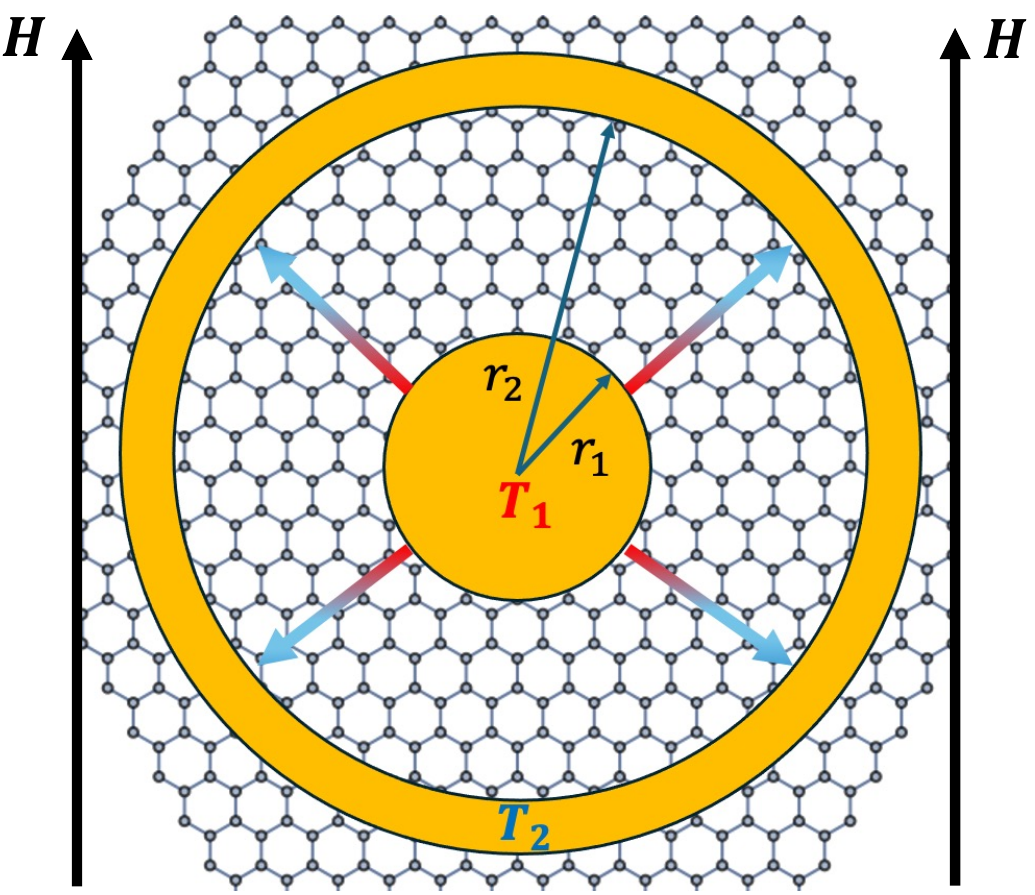}
\caption{A schematic for the graphene Corbino disk. Inner electrode has radius $r_1$ and is kept at temperature $T_1$. The outer electrode has radius $r_2$ and temperature $T_2<T_1$. 
The arrows pointing outwards  from the center mark heat fluxes carried by electron fluid in graphene layer which is subjected to the in-plane magnetic field $H$.}
\label{fig:Corbino}
\end{figure}

To explore this physics, a particularly useful device geometry is a graphene Corbino membrane, as illustrated in Fig. \ref{fig:Corbino}. In this setup, the continuity of currents enforces the electronic flow profile to be spatially nonuniform, even at uniform particle and entropy density. This amplifies the interplay of dissipative viscous, spin-diffusion, and spin-thermal effects. It should be noted that charge and thermal transport measurements in gated graphene Corbino devices have been reported, including observations of viscous and magneto-thermal effects \cite{Yan:2010,Faugeras:2010,Waissman:2024,Talanov:2024,Zeng:2024}. Additionally, powerful thermal and electrochemical potential imaging techniques can be implemented \cite{Waissman:2022,Ilani:2022,Brar:2023,Zeldov:2024}. Thus far, theoretical works have focused on the orbital effects of the magnetic field in the magnetotransport \cite{Falkovich:2019,Varlamov:2020,AL:2022,Gall:2023}, while hydrodynamic effects dominated by spin thermotransport remain unexplored, which motivates this work.

\section{Hydrodynamic theory}

To this end, consider a two-dimensional electron system subjected to an in-plane magnetic field. In the regime where the rate of momentum-conserving electron-electron collisions exceeds the rates of momentum, energy, and spin relaxation, the corresponding macroscopic hydrodynamic equations of the electron fluid can be written as follows: 
\begin{subequations}\label{eq:dot-npss}
\begin{equation}\label{eq:dot-n}
\partial_t n+\bm{\nabla}\cdot \bm{j}=0,
\end{equation}
\begin{equation}\label{eq:dot-s}
\partial_t s+\bm{\nabla}\cdot \bm{j}_s=\dot{s}, 
\end{equation}
\begin{equation}\label{eq:dot-sigma}
\partial_t\sigma+\bm{\nabla}\cdot \bm{j}_\sigma=\dot{\sigma},
\end{equation}
\begin{equation}\label{eq:dot-p}
\partial_t\bm{p}+k\bm{u}+\bm{\nabla}P+en\bm{\nabla}\phi=\bm{\nabla}\cdot\hat{\bm{\Sigma}}.
\end{equation}
\end{subequations}
Here $P$ is the pressure, $\bm{u}$ is the hydrodynamic velocity, $k$ denotes the disorder-induced friction coefficient, and $\hat{\bm{\Sigma}}\equiv\Sigma_{ij}$ denotes the viscous stress tensor \cite{LL-V6}
\begin{equation}\label{eq:stress}
\Sigma_{ij}=\eta(\partial_iu_j+\partial_j u_i)+(\zeta-\eta)\delta_{ij}\partial_k u_k, 
\end{equation}
with $\eta$ and $\zeta$ being, respectively, shear (first) and bulk (second) viscosity of the electron liquid. The electric potential $\phi$ is related to the electron charge density $en$ by the Poisson equation.
The densities of particles, entropy, momentum, and spin are denoted by $n$, $s$, $\bm{p}$, and $\sigma$ respectively. The corresponding current densities of particles, entropy, and spin are labeled respectively as $\bm{j}$, $\bm{j}_s$, and $\bm{j}_\sigma$.  The local rate of entropy production due to electron-electron collisions is denoted by $\dot{s}$ in Eq. \eqref{eq:dot-s} Likewise, $\dot{\sigma}$ in Eq. \eqref{eq:dot-sigma}  corresponds to the spin relaxation.  
The continuity equation for the spin current is written in analogy to the hydrodynamic theory of spin waves \cite{HH:1969}. It should be noted that only spin component along the field appears in this framework 
as the other spin components are not conserved due to spin precession.

We proceed to determine the entropy production rate. For this purpose, we recall that the entropy density $s$ is a function of the conserved quantities, and its differential is given by the general thermodynamic relation \cite{LL-V5}
\begin{equation}
ds=\frac{d\varepsilon}{T}-\frac{(\mu+e\phi)dn}{T}-\frac{\mu_\sigma d\sigma}{T}-\frac{\bm{u}\cdot d\bm{p}}{T},
\end{equation}
where $\mu$ and $\mu_\sigma$ are chemical potentials for particle and spin. Noting that all these thermodynamic variables correspond to conserved quantities, we express their time
derivatives in terms of the divergences of the corresponding currents from Eq. \eqref{eq:dot-npss}. This yields the following expression for $\dot{S}=\int \dot{s}d\bm{r}$,
\begin{align}
\dot{S}=\int\frac{1}{T}\left[-\bm{\nabla}\cdot\bm{j}_\varepsilon+(\mu+e\phi)\bm{\nabla}\cdot\bm{j}+\mu_\sigma\bm{\nabla}\cdot\bm{j}_\sigma\nonumber \right. \\ 
\left. +k\bm{u}^2+n\bm{u}\bm{\nabla}e\phi+\bm{u}\cdot\bm{\nabla}\cdot\hat{\bm{\Pi}} \right]d\bm{r},
\end{align}
where we used $\partial_t\varepsilon=-\bm{\nabla}\cdot\bm{j}_\varepsilon$ for the energy current density $\bm{j}_\varepsilon$, and similarly for $\partial_tn$, $\partial_t\sigma$, and $\partial_t\bm{p}$ from Eqs. \eqref{eq:dot-n}-\eqref{eq:dot-sigma}. The tensor $\hat{\bm{\Pi}}\equiv\Pi_{ij}=P\delta_{ij}-\Sigma_{ij}$ denotes the local part of the momentum flux.  Next, using the thermodynamic identity $\bm{\nabla}P=n\bm{\nabla}\mu+s\bm{\nabla}T+\sigma\bm{\nabla}\mu_\sigma$ and integrating by parts we get 
\begin{align}\label{eq:TdotS}
T\dot{S}=\int\left[-\bm{j}'_s\bm{\nabla}T-\bm{j}'\bm{\nabla}(\mu+e\phi)-\bm{j}'_\sigma\bm{\nabla}\mu_\sigma\nonumber \right. \\ 
\left. +k\bm{u}^2+\Sigma_{ij}\partial_ju_i\right]d\bm{r}
\end{align} 
where we introduced dissipative fluxes $\bm{j}'_s=\bm{j}_s-s\bm{u}$, $\bm{j}'=\bm{j}-n\bm{u}$, and $\bm{j}'_\sigma=\bm{j}_\sigma-\sigma\bm{u}$ that are proportional to gradients of the equilibrium parameters. The entropy flux is related to the energy flux as follows $\bm{j}_s=\frac{1}{T}[\bm{j}_\varepsilon-(\mu+e\phi)\bm{j}-\mu_\sigma\bm{j}_\sigma]$. 

In a situation with a vanishing electromotive force, $e\bm{\mathcal{E}}=-\bm{\nabla}(\mu+e\phi)$, the second term in the above expression is absent since $\bm{j}'=0$. 
Recalling the definition $\dot{S}=-\sum_iX_i\dot{x}_i$ from Ref. \cite{LL-V5} and identifying $\dot{x}_i$ with the entropy current $\bm{j}'_s$ and the spin diffusion current $\bm{j}'_\sigma$ we conclude that the role of the corresponding thermodynamic forces is played by $\bm{\nabla} T/T$ and $\bm{\nabla}\mu_\sigma/T$ respectively.   
Introducing the kinetic coefficients following the standard convention $\dot{x}_i=-\gamma_{ij}X_j$ we write
\begin{equation}
\left(\begin{array}{c}
\bm{j}'_s \\
\bm{j}'_\sigma
\end{array}
\right)=-\frac{1}{T}\left(\begin{array}{cc}\gamma_{11} & \gamma_{12} \\ \gamma_{21} & \gamma_{22} \end{array}\right)
\left(\begin{array}{c}
\bm{\nabla}T \\
\bm{\nabla}\mu_\sigma
\end{array}
\right).
\end{equation}
We can express $\gamma_{11}=\kappa$ in terms of the thermal conductivity $\kappa$, and $\gamma_{22}$ in terms the spin diffusion $D_\sigma$ as $\gamma_{22} =TD_\sigma$. 
Since the spin density changes sign with respect to time reversal symmetry and energy does not, the Onsager symmetry principle gives 
$\gamma_{12}(H)=-\gamma_{21}(-H)$. On the other hand both $\gamma_{12}(H)$ and $\gamma_{21}(H)$ are odd functions of the magnetic field $H$. 
Therefore, we conclude that $\gamma_{12}(H)=\gamma_{21}(H)$. Thus the matrix of kinetic coefficients is symmetric. 
In the following we denote $\gamma_{12}\equiv\gamma_\sigma$ and introduce 
\begin{equation}
\hat{\Xi}\equiv\hat{\gamma}/T=\left(\begin{array}{cc}
\kappa/T & \gamma_\sigma/T \\ \gamma_\sigma/T & D_\sigma
\end{array}\right). 
\end{equation}
As a reminder, for temperatures $T<E_F$ in the Fermi liquid theory $\kappa\sim E^2_F/T$ \cite{AK:1959} and $D_\sigma\sim A/T^2$ \cite{Hone:1961}, modulo logarithmic renormalizations specific to the 2D case, see e.g. Ref. \cite{Lyakhov:2003}. A detailed analytical calculations of these coefficients in the limit of weak interactions, $r_s=V/E_F\lesssim1$, where $V$ is the characteristic energy of Coulomb interaction 
and $E_F$ is the bare Fermi energy, can be found in Ref. \cite{SykesBrooker:1970}. 

For compactness of further expressions, it is convenient to introduce the column-vector notations 
\begin{equation}
\vec{x}=\left(\begin{array}{c} s \\ \sigma \end{array}\right),\quad 
\vec{\bm{X}}=\left(\begin{array}{c} \bm{\nabla}T \\ \bm{\nabla}\mu_\sigma\end{array}\right).
\end{equation}
Then it can be readily checked that the sum of the first and third term in the right hand side of Eq. \eqref{eq:TdotS} can be rewritten as $\vec{\bm{X}}^{\mathbb{T}}\hat{\Xi}\vec{\bm{X}}$, 
with the superscript $\mathbb{T}$ denoting the transposition. As a result, 
\begin{equation}\label{eq:dotS}
T\dot{S}=\int \left[\Sigma_{ij}\partial_ju_i+\vec{\bm{X}}^{\mathbb{T}}\hat{\Xi}\vec{\bm{X}}+k\bm{u}^2\right]d\bm{r},
\end{equation}
where summation over the repeated indices is implicit. The current densities take the final form in explicit notations 
\begin{subequations}\label{eq:currents}
\begin{equation}
\bm{j}_s=s\bm{u}-\frac{\kappa}{T}\bm{\nabla}T-\frac{\gamma_\sigma}{T}\bm{\nabla}\mu_\sigma,
\end{equation}
\begin{equation}
\bm{j}_\sigma=\sigma\bm{u}-\frac{\gamma_\sigma}{T}\bm{\nabla}T-D_\sigma\bm{\nabla}\mu_\sigma,
\end{equation}
\end{subequations}
where the first term in the right hand side represents the equilibrium components of the currents, while the remaining two terms in each expression represent the dissipative components. 
With the appropriate boundary conditions for a given device geometry, the system of equations Eqs. \eqref{eq:dot-npss}, \eqref{eq:dotS}, and \eqref{eq:currents} enable determination of the linear response transport coefficients (see Appendix \ref{app:A} for additional discussion). 

\section{Graphene Corbino device}

We apply the above formalism to consider heat transport in a graphene flake shaped like a Corbino disk, see Fig. \ref{fig:Corbino}. The aspect ratio of the disk is defined by the ratio of its radii, $p=r_2/r_1>1$.
In the linear response regime, a small heat current is induced in the device by a temperature difference $\Delta T=T_1-T_2$ between the electrodes. The calculations presented below will focus on the regime of charge neutrality, where $n\to0$. For simplicity, we also ignore the effect of bulk viscosity, which is known to vanish for both parabolic and linear spectra of quasiparticle excitations \cite{LL-V10}. The latter corresponds to the case of graphene monolayer.

In a steady state, the continuity equation of entropy and spin currents can be presented in the form specific to the cylindrical geometry \cite{Li:2022,AL:2022}
\begin{equation}\label{eq:xX}
\vec{x}u(r)-\hat{\Xi}\vec{X}(r)=\frac{\vec{I}}{2\pi r}, \quad \vec{I}=\left(\begin{array}{c}I_s \\ I_\sigma \end{array}\right), 
\end{equation}
where $u(r)$ is the radial hydrodynamic velocity and $\vec{X}(r)$ is the column-vector of thermodynamically conjugate forces, with $r\in [r_1, r_2]$ being the radial coordinate. 
The force balance condition exerted on an element of the fluid takes the form of the Navier-Stokes equation, which can be found with the help of Eqs. \eqref{eq:dot-p} and \eqref{eq:stress}: 
\begin{equation}\label{eq:NS}
\eta\hat{\Delta}u(r)-ku(r)=\vec{x}^{\mathbb{T}}\vec{X},\quad \hat{\Delta}=\frac{1}{r}\frac{d}{dr}\left(r\frac{d}{dr}\right)-\frac{1}{r^2}.
\end{equation}
Equations \eqref{eq:xX} and \eqref{eq:NS} determine the spatial dependence of the flow velocity, temperature gradients, and gradients of the spin chemical potential in
the interior of the Corbino disk in terms of the entropy and spin currents $\vec{I}$. The crucial point to realize here is that even though thermal resistance is defined under the global condition of vanishing spin current, 
locally dissipative parts of these currents are nonzero due to finite $\vec{X}(r)$ in the bulk of the flow. This can be seen explicitly. 

For this purpose, we use Eq. \eqref{eq:xX} to exclude $\vec{X}(r)$ and arrive at the equation for $u(r)$ in the form 
\begin{equation}\label{eq:NS-u}
\hat{\Delta}u-ku-\frac{u}{l^2}=-\frac{1}{2\pi r\eta}(\vec{x}^{\mathbb{T}}\hat{\Xi}^{-1}\vec{I}).
\end{equation}   
Here we introduced the characteristic length scale $l$ defined by
\begin{equation}
l^{-2}=\frac{\vec{x}^{\mathbb{T}}\hat{\Xi}^{-1}\vec{x}}{\eta}=\frac{s^2\left[D_\sigma-2\frac{\sigma\gamma_\sigma}{sT}+\frac{\sigma^2\kappa}{s^2T}\right]}{\eta\left[\frac{\kappa D_\sigma}{T}-\frac{\gamma^2_\sigma}{T^2}\right]}.
\end{equation}
The general solution of Eq. \eqref{eq:NS-u} consists of the linear superposition of solutions to the homogeneous equation and a particular solution due to the right hand side.  
The former is given by a linear combination of modified Bessel functions of the first and second kinds. These functions describe deviations of the hydrodynamic flow from that in the bulk of the disk.
Due to properties of Bessel functions they are exponentially localized near the inner and outer boundaries on the length scale of the order of $l$. For the graphene monolayer this length scale is of the order of thermal de Broglie length $l\sim l_T=v/T$, which in itself is of the order of electron mean free path $l_{\text{ee}}$. Therefore, these solutions contribute to the resistance of the contacts. Consideration of these effects goes beyond the limit of hydrodynamic theory. We are interested in the solution that describes the hydrodynamic mode in the bulk of the device. This solution can be easily found as a particular solution of the inhomogeneous equation, yielding the result
\begin{equation}\label{eq:ur}
u(r)=\frac{f}{2\pi r}\frac{\vec{x}^{\mathbb{T}}\hat{\Xi}^{-1}\vec{I}}{\vec{x}^{\mathbb{T}}\hat{\Xi}^{-1}\vec{x}},\quad f=\frac{1}{1+kl^2/\eta}.
\end{equation}
The factor defined by $f$ can be safely set to unity. Indeed, viscosity of the Dirac fluid is $\eta\sim (T/v)^2\sim l^{-2}_T$ \cite{Muller:2009}. In addition, friction coefficient can be related to the spatial average of the local density fluctuations $\delta n(\bm{r})$ of the globally charge neutral system, namely $k\sim\frac{e^2}{\varsigma}\langle\delta n^2\rangle$ \cite{Li:2020}, where $\varsigma$ is the intrinsic conductivity of graphene \cite{Mishchenko:2008,Fritz:2008,Kashuba:2008}. Therefore, $kl^2/\eta\sim \langle\delta n^2\rangle l^4_T\ll1$. Based on this argument, we set $f\to1$ in what follows. Knowledge of $u(r)$ enables us to compute dissipative friction term and also the viscous term in Eq. \eqref{eq:dotS}. For a purely radial flow, there are only two nonvanishing components of the stress tensor that contribute to the dissipated power. In cylindrical coordinates, these are
\begin{equation}\label{eq:Sigma}
\Sigma_{rr}=2\eta\frac{\partial u}{\partial r},\quad \Sigma_{\phi\phi}=2\eta\frac{u}{r}. 
\end{equation}

As the next step, we need to determine $\vec{X}(r)$, which governs contribution of the relative transport mode to the dissipation rate in Eq. \eqref{eq:dotS}. The required solution can be obtained by multiplying Eq. \eqref{eq:xX} by a row vector $\vec{\varkappa}^{\mathbb{T}}= (\sigma,-s)$ from the left. The result is 
\begin{equation}\label{eq:Xr}
\vec{X}(r)=-\frac{\vec{\varkappa}}{2\pi r}\frac{\vec{\varkappa}^{\mathbb{T}}\vec{I}}{\vec{\varkappa}^{\mathbb{T}}\hat{\Xi}\vec{\varkappa}},
\end{equation}
which completes solution of the hydrodynamic equations. 

\section{Spin-caloric resistance}

We introduce spin caloric resistance matrix $\hat{\mathcal{R}}$ that can be determined by equating the Joule heat, 
\begin{equation}\label{eq:P}
\mathcal{P}=\vec{\mathcal{I}}^{\mathbb{T}}\hat{\mathcal{R}}\vec{\mathcal{I}},\quad \vec{\mathcal{I}}^{\mathbb{T}}=(I_q, I_\sigma),
\end{equation}
to the rate of energy dissipation in the bulk flow as defined by Eq. \eqref{eq:dotS}. Here $I_q=TI_s$ is the heat current. The thermal resistance $R_{\text{th}}$ can be found by setting the spin current to zero $I_\sigma\to0$. The computed dissipation in Eq. \eqref{eq:P} is then simply $\mathcal{P}=\mathcal{R}_{11}I^2_q$. On the other hand, employing general thermodynamic relations, one finds the entropy production rate $\dot{S}=I_q\Delta T/T^2=R_{\text{th}}I^2_q/T^2$. As a consequence, from the condition $\mathcal{P}=T\dot{S}$ we determine that 
\begin{equation}\label{eq:Rth-def}
R_{\text{th}}=T\mathcal{R}_{11}=\frac{1}{TI^2_s}(T\dot{S})_{I_\sigma\to0}.
\end{equation}  
We insert Eqs. \eqref{eq:ur}, \eqref{eq:Sigma}, and \eqref{eq:Xr} into Eq. \eqref{eq:dotS}, set $I_\sigma\to0$, compute each of the radial integrals individually, and bring the resulting expression into Eq. \eqref{eq:Rth-def} that will cancel $I^2_s$ in the denominator. The resulting formula can be written as a sum of respective contributions (see Appendix \ref{app:B} for technical details)
\begin{equation}\label{eq:Rth}
R_{\text{th}}=R_\Sigma+R_\Xi+R_k.
\end{equation}
The first term describes viscous dissipation modified by the spin diffusion and spin thermocurrent, which reads 
\begin{subequations}\label{eq:Rth-sum}
\begin{equation}\label{eq:R-Sigma}
R_\Sigma=\frac{\eta(p^2-1)}{\pi Ts^2r^2_2}\frac{\left(D_\sigma-\frac{\sigma\gamma_\sigma}{sT}\right)^2}{\left(D_\sigma+\frac{\sigma}{s}\beta_\sigma\right)^2},
\end{equation}
with the notation $\beta_\sigma=(\sigma\kappa/sT)-(2\gamma_\sigma/T)$. The second term captures contribution of the relative mode and is expressed solely in terms of the spin-dependent intrinsic dissipative coefficients of matrix $\hat{\Xi}$ 
 \begin{equation}
R_\Xi=\frac{1}{2\pi T}\frac{(\sigma/s)^2\ln p}{D_\sigma+\frac{\sigma}{s}\beta_\sigma}.
\end{equation}
The third term describes an extra contribution due to disorder-induced friction 
\begin{equation}
R_k=\frac{k\ln p}{2\pi Ts^2}\frac{\left(D_\sigma-\frac{\sigma\gamma_\sigma}{sT}\right)^2}{\left(D_\sigma+\frac{\sigma}{s}\beta_\sigma\right)^2}.
\end{equation}
\end{subequations}
Eqs. \eqref{eq:Rth} and \eqref{eq:Rth-sum} represent the main results of this work. Next we analyze several limiting cases. 

In pristine systems, $k\to0$, and without magnetic field, $\sigma\to0$, the result simplifies to $R_{\text{th}}=\eta(p^2-1)/\pi Ts^2r^2_2$ (see Appendix \ref{app:C} for an alternative derivation). For a graphene monolayer $\eta\sim s\sim (T/v)^2$, so that $R_{\text{th}}\propto 1/T^3$. In contrast, in bilayer graphene (BLG) the entropy density has different temperature dependence, $s\sim m^* T$, where $m^*$ is the effective mass of the band structure.  From the bound on the viscosity to entropy ratio, known from the strongly interacting quantum field theories \cite{Son:PRL05}, one can deduce that $\eta\propto T$, therefore $R_{\text{th}}\propto 1/T^2$ in BLG. These predictions can be tested in experiments.  

The leading field dependent correction to $R_{\text{th}}$ comes from $R_\Xi$. 
At low magnetic fields, we write the spin density in the form $\sigma=\chi H$, where $\chi$ denotes the spin susceptibility. 
As a result, for the relative thermal magnetoresistance, one obtains 
\begin{equation}\label{eq:relative-MR} 
\Delta R_{\text{th}}(H)\equiv\frac{R_{\text{th}}(H)-R_{\text{th}}(0)}{R_{\text{th}}(0)}\approx\frac{p^2\ln p}{2(p^2-1)}\frac{r^2_1(\chi H)^2}{\eta D_\sigma},
\end{equation} 
which is controlled by the spin diffusion coefficient. We conclude that the spin caloric effect enables the extraction of $D_\sigma$ and, thus, the spin conductance via the corresponding Einstein relation, in the transport regime dominated by electron collisions. This approach could offer valuable complementary experimental tools in addition to earlier methods based on spin Coulomb drag measurements \cite{Weber:2005}.

These results can be readily generalized to the magnetotransport away from charge neutrality. There are three key modifications that need to be incorporated. (i) At finite particle density $n$, the thermoelectric effects couple charge and heat flows. Therefore, we have to augment Eq. \eqref{eq:currents} by the respective expression for the electric current density 
\begin{equation}
\bm{j}_e=en\bm{u}+\varsigma\bm{E}-\frac{\gamma}{T}\bm{\nabla}T,
\end{equation} 
where $\gamma$ is the intrinsic thermoelectric coefficient and the electromotive force (EMF) $e\bm{E}$ is the gradient of the local electrochemical potential. It is important to realize that finite $\bm{E}$ is generated even in a purely thermal bias setup. Since thermal resistance is defined at vanishing spin and electric currents, vanishing of $\bm{j_e}$ implies a local relationship $\bm{E}=-(en/\varsigma)\bm{u}$.   
Note that at $n\ll s$ the thermoelectric contribution to $\bm{j}_e$ may be neglected because $\gamma/T\sim n/s\ll1$. (ii) The Coulomb force $-en\bm{E}$ must be added to the right-hand-side of the force balance condition, Eq. \eqref{eq:NS}. Inserting $\bm{E}$ into the force-balance condition gives a force term $(en)^2\bm{u}/\varsigma$, which amounts to the following redefinition of the effective friction coefficient in Eq. \eqref{eq:NS}, $k\to k(n)=k+\frac{e^2}{\varsigma}n^2$. The form of the equation remains intact thus its solution for $\bm{u}$, which is still given by Eq. \eqref{eq:ur}. (iii) An additional term must be added to the entropy production rate in Eq. \eqref{eq:dotS}, which takes a simple form
\begin{equation}\label{eq:dotS-E}
T\dot{S}=\varsigma\int\bm{E}^2d\bm{r}
\end{equation} 
with the neglect of thermoelectric contribution $\propto\gamma/T\ll1$. With these ingredients, we use Eq. \eqref{eq:ur} to calculate Eq. \eqref{eq:dotS-E} and thus determine additional contribution to thermal resistance due to intrinsic electric conductivity 
\begin{equation}
R_\varsigma=\frac{e^2}{\varsigma}\frac{n^2}{s^2}\frac{f^2\ln p}{2\pi T}\frac{\left(D_\sigma-\frac{\sigma\gamma_\sigma}{sT}\right)^2}{\left(D_\sigma+\frac{\sigma}{s}\beta_\sigma\right)^2}.
\end{equation}
In this limit $f$ defined in Eq. \eqref{eq:ur} has noticeable density dependence. Therefore $R_\Sigma$ and $R_k$ should be also multiplied by a factor $f^2$, since $u\propto f$, but not $R_\Xi$, which remains the same. The limit of charge neutrality corresponds to $n\ll\sqrt{\langle\delta n^2\rangle}$. In modern encapsulated devices density inhomogeneity of monolayer graphene is in the range $\sqrt{\langle\delta n^2\rangle}\sim (5\div10)\times 10^9$ cm$^{-2}$. This is still well within the domain of Dirac plasma $\{n,\sqrt{\langle\delta n^2\rangle}  \}<s$ at $T\sim 100$ K. In this limit one may approximate $f\approx s^2/(s^2+\frac{e^2}{\varsigma}n^2)$. Therefore, $R_\Xi$ remains the leading spin-dependent contribution to thermal resistance at all relevant densities $n<s$.    

\section{Estimates}

It is instructive to estimate the order of magnitude of computed thermal resistances. For this purpose consider Eq. \eqref{eq:Rth} in the limit $\sigma\to0$ as an example. 
Recall that for a monolayer graphene $\eta\sim s\sim(k_BT/\hbar v)^2$, where Boltzmann and Planck constants were restored to have proper units in SI system. Let us take $T\sim 100$ K and $v\sim 10^6$ m/s. The typical radius of the Corbino disk in experiments is $r_{1,2}\gtrsim 1\mu$m.  For this values, let us convert the conventional K/W units of $R_{\text{th}}$ into $\Omega$ units of electrical resistance by using the Lorenz number $L$ as a conversion factor through the Wiedemann-Franz law $L=R_{\text{el}}/TR_{\text{th}}$. This gives $R_{\text{th}}\sim (\hbar/e^2)1/(sr^2)\sim 100\, \Omega$. The measured values of thermal resistance reported in Ref. \cite{Talanov:2024} are in the range $\sim 50\div600\, \Omega$ for temperatures in the range $T\sim 25\div200$ K. The estimate matches well. 

The kinematic viscosity measured in Corbino geometry as reported in Refs. \cite{Talanov:2024,Zeng:2024} is in the range $\nu\sim0.1\div0.5$ m$^2$/s. The kinematic viscosity is given by the shear viscosity per mass density $\nu=\eta/\rho$. The mass density of the Dirac thermal cloud can be estimated as $\rho\sim T^3/v^4$. Putting this together and restoring proper conversion units one finds $\nu\sim\hbar v^2/k_BT$, which gives $\nu\sim0.1$ m$^2$/s for $T=100$ K. Again this gives  good agreement. 

It is harder to estimate the magnitude of the relative MR in Eq. \eqref{eq:relative-MR} because of unknown values of spin diffusion and spin susceptibility in the hydrodynamic regime, but form the structure of the expression it is expected that the result is significant even at relatively weak field.  

\section*{Acknowledgments}

I am grateful to Anton Andreev and Philip Kim for insightful discussions. This work was supported by the U.S. Department of Energy (DOE), Office of Science, Basic Energy Sciences (BES) under Award No. DE-SC0020313 and H. I. Romnes Faculty Fellowship provided by the University of Wisconsin-Madison Office of the Vice Chancellor for Research and Graduate Education with funding from the Wisconsin Alumni Research Foundation.

\appendix 

\section{Assumptions and approximations}\label{app:A}

In the main part of this paper, the calculation of the spin thermoresistance is described under the assumption of the hydrodynamic regime of electron flow, characterized by frequent collisions and strong correlations. The key physical insight developed in this work relates to the pronounced spin-heat transport coupling near charge neutrality, where thermal anomalies are expected to be strong. The coupling equations for the spin and entropy currents in the presence of a temperature gradient and a spin chemical potential were derived starting from the generic conservation equations for entropy, charge, spin, and the force-balance condition. These equations are solved in a Corbino disk geometry.

The applicability of the hydrodynamic limit assumes that the electron mean free path $l_{\text{ee}}$, due to momentum-conserving collisions, is the shortest length scale in the problem. For monolayer graphene (MLG), this scale is of the order of the thermal de Broglie length $l_{\text{ee}}\sim l_T=v/T$. Momentum-relaxing collisions on disorder induce friction on the liquid flow. For smooth disorder potential, the friction coefficient $k$ can be expressed in terms of the local variations of the particle density $\delta n(\bm{r})$ and entropy density $\delta s(\bm{r})$ as follows \cite{Li:2020} 
\begin{equation}
k=\frac{\langle(s\delta n-n\delta s)^2\rangle}{2\left(\frac{n^2\kappa}{T}-\frac{2ns\gamma}{T}+\frac{s^2\varsigma}{e^2}\right)}.
\end{equation}
At charge neutrality, $n\to0$, intrinsic thermoelectric coefficients vanishes, $\gamma\to0$, and $k$ simplifies to $k=\frac{e^2}{2\varsigma}\langle\delta n^2\rangle$, which is determined by the intrinsic conductivity. For MLG devices the intrinsic conductivity is known to be
of the order of conductance quantum $\sim e^2/2\pi$ (in units of $\hbar=1$) modulo logarithmic renormalizations in the weak-coupling
theory \cite{Mishchenko:2008,Fritz:2008,Kashuba:2008}
\begin{equation}
\varsigma=\frac{e^2}{2\pi\alpha^2_T},\qquad \alpha_T=\frac{\alpha_g}{1+\frac{\alpha_g}{4}\ln\frac{\Lambda}{T}},
\end{equation} 
where $\alpha_g=e^2/\varepsilon v$ is the temperature-independent dimensionless interaction constant ($\varepsilon$ is the dielectric constant of the surrounding medium), which determines the bare strength of the electron-electron interactions, and $\Lambda$ is the cutoff in the scheme of the renormalization group.

In the model explored in this paper, a long-range disorder potential is assumed, characterized by a correlation radius $\xi$ longer than the equilibration length but shorter than the distance between the electrodes, namely $l_T< \xi < r_2-r_1$. These conditions can be securely met in modern devices. Indeed, for boron nitride encapsulated graphene devices, scanning probes reveal that the correlation radius of these fluctuations is somewhere in the range $\xi\sim 100$ nm and local strength is in the range of $\sim 5$ meV. Corbino devices can be designed with the radial difference $\sim 5 \mu$m.  

In the model long spin relaxation time is assumed. This is particularly relevant to graphene multilayers, as the spin relaxation, even at room temperature, in the range of nanoseconds, with corresponding spin diffusion lengths reaching $l_\sigma\sim 10 \mu$m. 

\section{Spin-caloric magnetoresistance}\label{app:B}

In this section we provide additional details for the computation of the thermal resistances $R_\Sigma$, $R_\Xi$, and $R_k$, which enter Eq. (20) in the main text.  

\subsection{Viscous contribution}

A particular solution of the hydrodynamic equations for the radial flow is just the consequence of the continuity of entropy current. As shown in the main text [Eq. (15)], it takes the form  
\begin{equation}
u(r)=\frac{A}{2\pi r},\quad A=\frac{(\vec{x}^{\mathbb{T}}\hat{\Xi}^{-1}\vec{I})}{(\vec{x}^{\mathbb{T}}\hat{\Xi}^{-1}\vec{x})}. 
\end{equation}
This expression gives us two nonvanishing components of the stress tensor 
\begin{equation}\label{eq:Sigma}
\Sigma_{rr}=2\eta\frac{\partial u}{\partial r}=-2\eta\frac{A}{2\pi r^2},\quad \Sigma_{\phi\phi}=2\eta\frac{u}{r}=2\eta\frac{A}{2\pi r^2}. 
\end{equation}
Using the definition of the stress tensor $\Sigma_{ij}$ \cite{LL-V6}, the entropy production can be equivalently rewritten as follows 
\begin{equation}
T\dot{S}=\frac{1}{2\eta}\sum_{ij}\int \Sigma^2_{ij}d^2r=\frac{1}{2\eta}\int (\Sigma^2_{rr}+\Sigma^2_{\phi\phi})d^2r
\end{equation} 
where integration goes over the two-dimensional area of the Corbino disk. Since $\Sigma^2_{rr}=\Sigma^2_{\phi\phi}$ for $u\propto 1/r$ flow, we thus have 
\begin{equation}
T\dot{S}=\frac{1}{2\eta}\times 2\times\int^{r_2}_{r_1}\left(\frac{\eta A}{\pi r^2}\right)^22\pi rdr=\frac{\eta A^2}{\pi}\left(\frac{1}{r^2_1}-\frac{1}{r^2_2}\right).
\end{equation} 
As the next step, we calculate the coefficient $A$, which consists of two terms:
\begin{subequations}
\begin{align}
&\vec{x}^{\mathbb{T}}\hat{\Xi}^{-1}\vec{I}=\frac{1}{\mathrm{Det}\,\hat{\Xi}}\left[\left(sD_\sigma-\sigma\frac{\gamma_\sigma}{T}\right)I_s+\left(\sigma\frac{\kappa}{T}-s\frac{\gamma_\sigma}{T}\right)I_\sigma\right], \\ 
&\vec{x}^{\mathbb{T}}\hat{\Xi}^{-1}\vec{x}=\frac{1}{\mathrm{Det}\,\hat{\Xi}}\left[s^2D_\sigma-2s\sigma\frac{\gamma_\sigma}{T}+\sigma^2\frac{\kappa}{T}\right].
\end{align}
\end{subequations} 
Therefore
\begin{equation}\label{eq:A}
(A)_{I_\sigma\to0}=\frac{I_s}{s}\frac{D_\sigma-\frac{\sigma\gamma_\sigma}{sT}}{D_\sigma+\frac{\sigma^2\kappa}{s^2T}-2\frac{\sigma\gamma_\sigma}{sT}}.
\end{equation}
The corresponding resistance is given by 
\begin{align}
R_\Sigma&=\frac{1}{TI^2_s}(T\dot{S})_{I_\sigma\to0}\nonumber \\ 
&=\frac{\eta}{\pi Ts^2}\left(\frac{1}{r^2_1}-\frac{1}{r^2_2}\right)
\left(\frac{D_\sigma-\frac{\sigma\gamma_\sigma}{sT}}{D_\sigma+\frac{\sigma^2\kappa}{s^2T}-2\frac{\sigma\gamma_\sigma}{sT}}\right)^2
\end{align}
After a trivial algebraic step, this expression can be reduced to the form of Eq. (21a) from the main text. 

\subsection{Spin diffusion contribution}

The solution for the spatial distribution of the thermodynamic forces in the bulk of the Corbino device is given by [see Eq. (17) in the main text]:
\begin{equation}
\vec{X}=-\frac{\vec{\varkappa}}{2\pi r}\frac{\vec{\varkappa}^{\mathbb{T}}\vec{I}}{\vec{\varkappa}^{\mathbb{T}}\hat{\Xi}\vec{\varkappa}}.
\end{equation}
In the entropy product rate we need an expression 
\begin{equation}
\vec{X}^{\mathbb{T}}\hat{\Xi}\vec{X}=\frac{B}{(2\pi r)^2},\quad B=\frac{(\vec{\varkappa}^{\mathbb{T}}\vec{I})^2}{\vec{\varkappa}^{\mathbb{T}}\hat{\Xi}\vec{\varkappa}}, 
\end{equation}
which gives us 
\begin{equation}
T\dot{S}=\int \vec{X}^{\mathbb{T}}\hat{\Xi}\vec{X} d^2r=\int^{r_2}_{r_1}\frac{B}{(2\pi r)^2}2\pi rdr=\frac{B}{2\pi}\ln\frac{r_2}{r_1}. 
\end{equation}
To obtain $B$ we calculate vector products 
\begin{subequations}
\begin{align}
&\vec{\varkappa}^{\mathbb{T}}\vec{I}=\sigma I_s-sI_\sigma, \\ 
&\vec{\varkappa}^{\mathbb{T}}\hat{\Xi}\vec{\varkappa}=s^2D_\sigma-2s\sigma\frac{\gamma_\sigma}{T}+\sigma^2\frac{\kappa}{T}.
\end{align}
\end{subequations}
As a result, for the corresponding resistance one obtains 
\begin{equation}
R_\Xi=\frac{(\sigma/s)^2\ln(r_2/r_1)}{2\pi T}\left(\frac{1}{D_\sigma+\frac{\sigma^2\kappa}{s^2T}-2\frac{\sigma\gamma_\sigma}{sT}}\right),
\end{equation}
where we used $(B)_{I_\sigma\to0}$, that reduces to Eq. (21b) from the main text of the paper.   

\subsection{Disorder friction contribution}

The dissipative power associated with the disorder-induced friction takes the form 
\begin{equation}
T\dot{S}=k\int u^2d^2r=k\int^{r_2}_{r_1}\left(\frac{A}{2\pi r}\right)^22\pi rdr=\frac{kA^2}{2\pi}\ln\frac{r_2}{r_1}
\end{equation}
The corresponding resistance is given by 
\begin{equation}
R_k=\frac{1}{TI^2_s}(T\dot{S})_{I_\sigma\to0}=\frac{k\ln(r_2/r_1)}{2\pi Ts^2}
\left(\frac{D_\sigma-\frac{\sigma\gamma_\sigma}{sT}}{D_\sigma+\frac{\sigma^2\kappa}{s^2T}-2\frac{\sigma\gamma_\sigma}{sT}}\right)^2,
\end{equation}
where we used $(A)_{I_\sigma\to0}$ from the previous section, Eq. \eqref{eq:A}. The above expression can be easily reduced to the form of Eq. (21c). 

\section{Thermal resistance}\label{app:C}

To elucidate the physical origin of the viscous contribution to the thermal resistance, it is instructive to derive it from an alternative consideration. 
To this end, observe one crucial point, that the obtained solution for the thermodynamic force $\vec{X}(r)$, obeys the following relation $\vec{x}^{\mathbb{T}}\vec{X}=0$. In explicit notations it reads 
\begin{equation}
s\bm{\nabla} T+\sigma\bm{\nabla}\mu_\sigma=0.
\end{equation}  
It can be understood as the force balance condition exerted on the element of the fluid where local forces due to the temperature gradients are compensated by forces on electron spins arising from the local gradients of the spin chemical potential. In total, this relation reveals the global expulsion of the force density from the bulk of the device. Consider now the flow without spin polarization $\sigma\to0$. This means that $\bm{\nabla}T=0$ so that the temperature of electron liquid is constant $T(r)=T_l$. Yet, to induce the heat flow, we must apply the temperature bias $\Delta T=T_1-T_2$ across the entire device between inner and outer electrodes, which is not consistent with the constant temperature profile. This apparent paradox is resolved by the fact that there are temperature drops at the contacts.  
These drops create an excess pressure difference that is compensated by the radial viscous stresses arising in the liquid that exert additional radial force on the contacts. Recall the thermodynamic identity for the pressure differential, $dP=nd\mu+sdT$ \cite{LL-V5}, where $\mu$ is the chemical potential. At charge neutrality, $n\to0$, the extra pressure is only due to temperature drops. Therefore, 
the pressure jumps at the boundary with the contacts are given by      
\begin{equation}\label{eq:P-drop}
s(T_i-T_l)=\Sigma_{rr}(r_i),
\end{equation}
where $T_i$ is the temperature of the $i$-th contact. Note that the temperature of the liquid, $T_l$ is either higher or lower than the temperatures of both leads, depending on the direction of the heat flow. This is a consequence of the bulk character of entropy production. It should be noted that a similar consequence of force expulsion occurs in charge transport away from charge neutrality. The voltage drop between the contacts and the electron liquid has the same sign at both boundaries. 

Taking the difference in Eq. \eqref{eq:P-drop} we obtain 
\begin{equation}
s(T_1-T_2)=\Sigma_{rr}(r_1)-\Sigma_{rr}(r_2)=-\frac{\eta A}{\pi}\left(\frac{1}{r^2_1}-\frac{1}{r^2_2}\right). 
\end{equation}
where we used Eq. \eqref{eq:Sigma}. Without spin polarization $A=I_s/s$, see Eq. \eqref{eq:A}, therefore 
\begin{equation}
s\Delta T=-\frac{\eta I_s}{\pi s}\left(\frac{1}{r^2_1}-\frac{1}{r^2_2}\right). 
\end{equation}
In the linear response to $\Delta T$, the heat current is given by $I_q=-G_{\text{th}}\Delta T$, where $G_{\text{th}}$ is thermal conductance. 
Using $I_q=TI_s$ one finds the corresponding resistance 
\begin{equation}\label{eq:Rth}
R_{\text{th}}\equiv G^{-1}_{\text{th}}=-\frac{\Delta T}{TI_s}=\frac{\eta}{\pi Ts^2}\left(\frac{1}{r^2_1}-\frac{1}{r^2_2}\right).
\end{equation}
It coincides with Eq. \eqref{eq:R-Sigma} at $\sigma\to0$, which was obtained by matching dissipated power to Joule heating. This argument can be generalized to the case of finite spin polarization.

\bibliography{biblio}

\end{document}